# Millikelvin-compatible apparatus for studies of quantum materials under uniaxial stress

*Donovan Davino[1*], Jacob Franklin[1*] & Ilya Sochnikov[1,2, a)]*

**\***These authors contributed equally to this work.

**AFFILIATIONS**

[1]*Physics Department, University of Connecticut, Storrs Connecticut, USA 06269*
[2]*Institute of Material Science, University of Connecticut, Storrs Connecticut, USA 06269*

[a)] *Author to whom correspondence should be addressed. Electronic Mail:*
*ilya.sochnikov@uconn.edu*

**Abstract**
Various new phenomena emerge in quantum materials under elastic deformations, such as hydrostatic or uniaxial stresses. In particular, using uniaxial strain or stress can help to tune or uncover specific structural or electronic orders in materials with multiple coexisting phases. Those phases may be associated with a quantum phase transition requiring a millikelvin environment combined with multiple experimental probes. Here, we describe our unique apparatus, which allows *in situ* tuning of strain in large samples inside a dilution refrigerator while the samples are monitored via an optical microscope. We describe the engineering details and show some typical results of characterizing superconducting strontium titanate under stress. This letter should serve as a practical reference for experts in ultra-low temperature experimental physics involving uniaxial stresses or strains.

**Introduction**

In a wide variety of quantum materials, strain plays an important role. For example, in strontium titanate, which is a low critical temperature ($T_c$) superconductor, questions about our fundamental understanding of electron lattice coupling and ways to enhance $T_c$ are at the forefront of the modern quest to understand quantum phase transitions.[1]–[7] In high temperature superconductors, pressure and strain may be the keys to potentially even higher temperature superconductivity.[8] For example, recent observations of the high-pressure-induced (compressive strain) record-breaking superconductivity in sulfur hydride at about 200 K are a vivid demonstration that strain and pressure effects are a promising control parameter to make significant advances in tuning superconducting transitions.[9] In magnetic systems, such as quantum liquid candidates and quantum spin-ices, strain and lattice effects may be the key in realizing new exotic quasiparticles.[10], [11] In topological quantum materials, the strain might be key to realizing dissipationless protected states that may be useful for future quantum computing approaches. [12]–[15] Since many of these effects happen at millikelvin temperatures, it is important to develop strain cells that work reliably at such low temperatures in dilution refrigerators. This work presents such a millikelvin apparatus with examples of strain effects detected via optical, magnetic and electronic means in compressed superconducting strontium titanate.

**Strain cell**

Figure 1 illustrates our thermally compensated copper strain cell. The cell has a copper lever that compresses or stretches the sample. The lever is controlled by a lead screw that is rotated by a high-torque, low-speed geared DC motor. The lead screw driving rod is fed through the entire height of the Bluefors LD-250 dilution refrigerator (Figure 2), with thermalization points at every cold plate, lead-bronze bearings at some low-temperature stages, and a high-vacuum ferrofluid rotational feedthrough at the room temperature stage.

The strain in the samples is measured with a miniature gauge directly attached to the sample. [2], [6], [7] The cell was designed to accommodate the changes in strontium titanate sample size upon cooling with accuracy of about 0.01% in residual strain. We verified the accuracy by comparing the gauge readings on clamped samples to the readings of the strain gauges mounted on free-standing samples (data not shown). We typically mounted the samples with attached gauges in the copper strain cell using Stycast epoxy, leaving ~5-7 mm of the sample length for the probing of superconducting properties and optical imaging. Therefore, as-cooled samples typically had very close to zero strain. [2] Thus, this design provides small strains and minimizes dislocation defects in the sample.

This performance metric of our strain device is comparable to that of commercial and non-commercial piezo-based devices, [16] but our copper clamp is essential for efficient thermalization at temperatures in a dilution refrigerator. We chose to develop an all-metal strain rig (lead screw clamping) due to better control of tensile strain and the ability to apply very large forces (>400 N). These forces translate into more than 1 GPa of stress on samples. The disadvantage which results from the larger size of our device is that it causes increased cooling times. However, this disadvantage is insubstantial, and the larger size also results in the advantage that our clamp allows us to attach resistive gauges directly to large samples for *in situ* monitoring of strain, as well as other probes such as optics, magnetic susceptibility and scanning SQUID (scanning SQUID data not shown). Even though we preferred to work with 4-10 mm long crystals when those were available, our clamp can accommodate samples that are about 1-20 mm long. The cross section of the samples should be determined according to the maximal desired strain or stress.

The simulation (finite element analysis in Fusion 360) shown in Figure 1 was used to analyze the expected experimental conditions of the strain cell. The real-world strain cell was made from H04 Copper with a yield strength of about 240 Megapascal. The cell is then pre-loaded so that it can stretch or compress the sample through force applied to the lever arm by the lead screw. The simulation shows that at a preloaded angle of 1.4 degrees (measured from the lever arm to the base) 400N should be applied to displace the sample location on the lever arm by .75mm. At this point the end of the arm has been displaced 2.2 mm making it an arc symmetrical relative to the sample. One benefit of preloading the sample like this is that it allows for a more uniform relationship when force is applied to the lever arm compared to starting the lever arm completely parallel to the beam. Additionally, because 400 N has already been applied to the lever arm, when force is taken away the sample can then be stretched (tensile stress). The simulation shows that at 400N, a safety factor of 2 is reached in the copper parts (plastic deformation occurs below 1). This shows that there is still some room to add more force to the sample which allows for further compression while still being able to compress the sample.

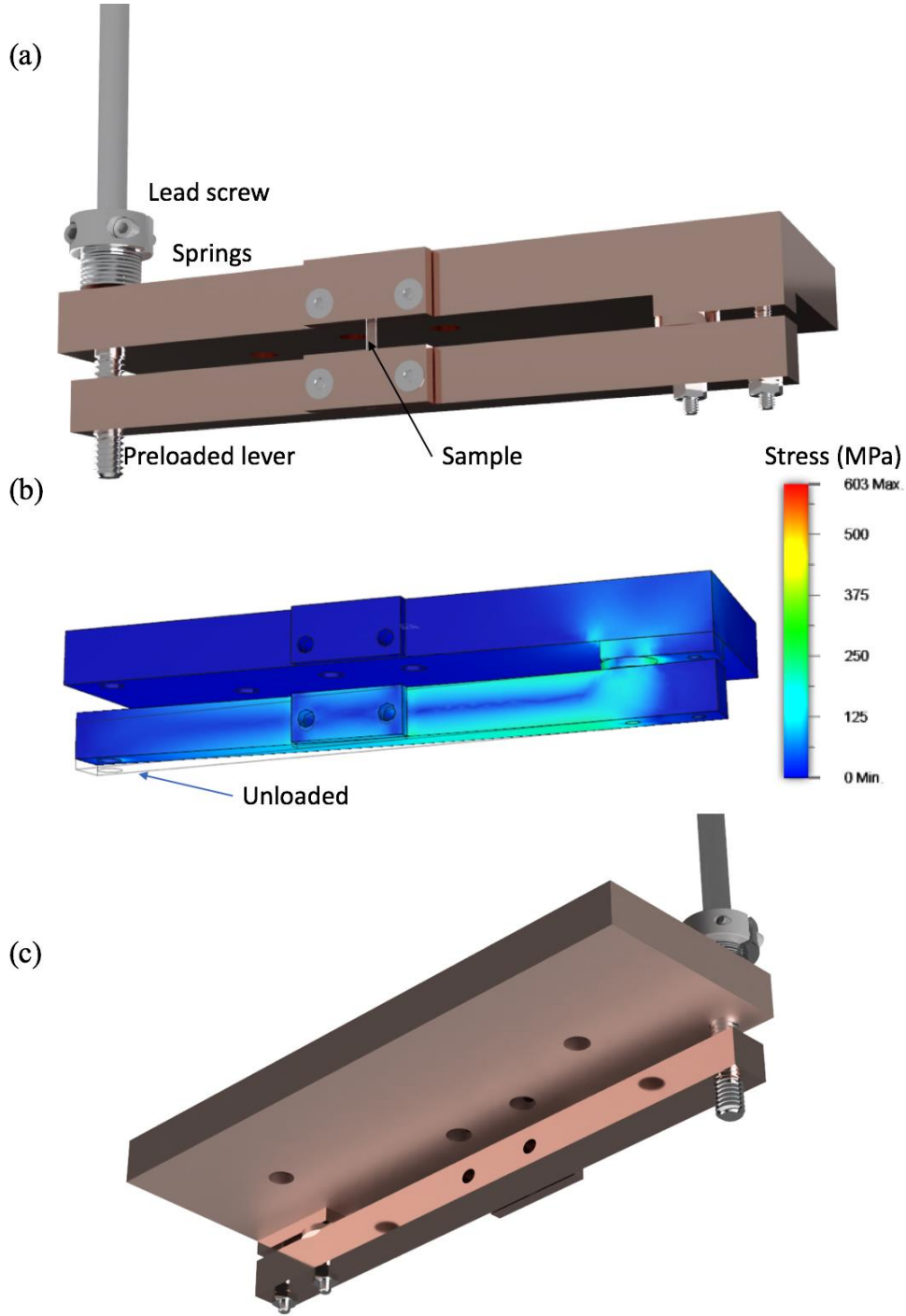

*Figure 1. **The copper strain compressive and tensile cell.** (a) The geometry of the clamp, combined with the copper and stainless-steel parts, compensates for the various thermal shrinkages of STO*

*samples and the adjacent copper parts. The main material is oxygen-free copper for use in a dilution refrigerator. A simple cell design is critical to the success of the measurements of samples that are essentially free of residual strain when cooled to temperatures of 10s of mK. (b) Finite element stress analysis of the cell in preloaded state in units of MPa. The lever is parallel to the baseplate in the preloaded state allowing for uniform stress in the sample. (c) A view of the clamp from under and behind it to assist in the visualization of its geometry. The cell is able to provide >400 N of compressive and tensile force. No commercial cells can achieve these values at such low temperatures to the best of our knowledge.*

**Optical monitoring of tetragonal domains in a dilution refrigerator**

In addition to the cell, we have optical, magnetic and electrical probes. The optical microscope is the most unique among the three and is described here.

We constructed a polarizing optical microscope in our BF-250 dilution refrigerator using mostly conventional optical components. We took care to mount the lenses and the mirrors in as strain-free a manner as possible. Some lenses simply rested on stopping rings inside aluminum lens tubes to avoid cracking due to thermal contraction. We degreased and removed plastic parts from Thorlabs (New Jersey, USA) and Newport (California, USA) optomechanical components to make them more compatible with vacuum and cryogenics. Otherwise, the microscope has a conventional geometry (Figure 2): with an objective and ocular lens and a beam splitter for illumination in between. An LED source (Thorlabs MCWHF2) is located outside the cryostat. A silica HV vacuum-compatible optical fiber 800 μm in diameter (Thorlabs MV63L1) carries the light. A polymer film with no glass layers polarizes the light and two additional fused silica lenses refocus it while passing it to a 50:50 beam splitter (Thorlabs BSW10) in a Köhler illumination scheme. We installed a series of infrared-opaque windows at each cold plate of the cryostat. The light focuses on the sample with a standard single AR-A 1'' diameter 250 mm focal length fused silica objective lens (with two adjustable rerouting mirrors), reflects back from the sample, passes through the same beam splitter, a standard AR-A fused silica 1" diameter 750 mm focal length condenser lens, passes through another standard polymer film polarizer, and then through a standard UHV fused silica AR-A viewport to a AmScope MU1403B 14MP High-Speed USB 3.0 CMOS camera. The camera acquired the images, while additional optomechanical components aligned the camera outside the cryostat. The total optical path is over one meter long, requiring very fine manual alignment. We obtained a spatial resolution of ~5-8 μm from the microscope.

Representative optical images of strontium titanate at a range of strains is shown in Figure 3. When the samples undergo the antiferrodistortive phase transition around 110 K, large domain regions form. We determined that samples can be easily detwinned under stress in the (110) direction. Figure 3 shows 0.5x1.5x10 mm$^3$ multi-domain vs. detwinned states of the sample. Further compression resulted in suppression of the critical temperature in this strontium titanate sample, as detected in the resistive and magnetic transition (Figure 3), all measured simultaneously. The changes in superconductivity are interpreted as being due to influence of strain on the phonon modes and the proximity to the ferroelectric quantum phase transition.[2], [6] Additional data from other samples and reproducibility of the results related to the sample mounting procedure are discussed elsewhere.[2], [6]

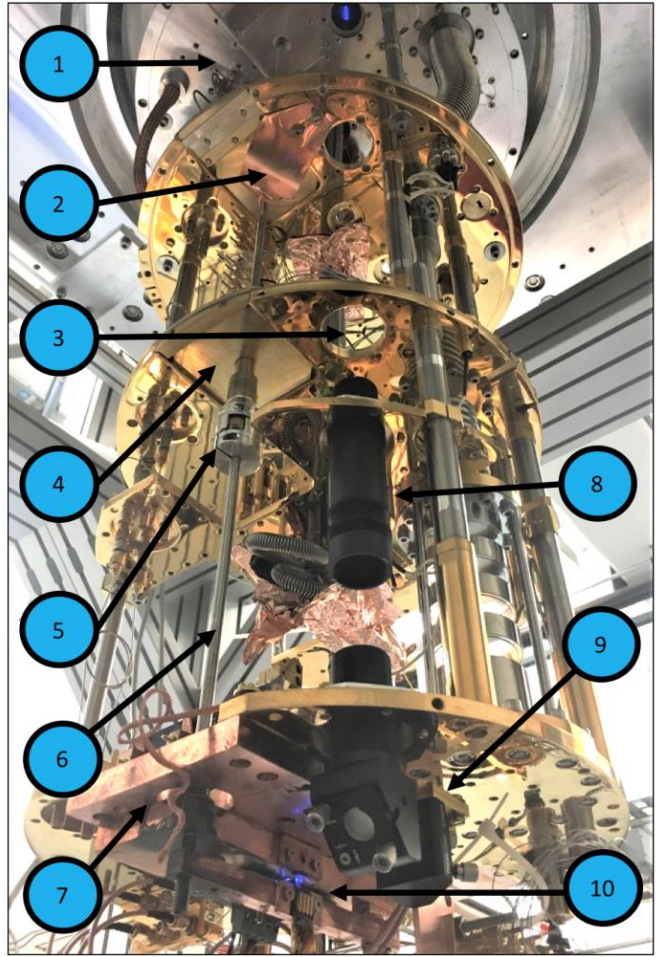

*Figure 2. **Polarized optical microscopy in the Bluefors dilution refrigerator to image the strained samples in situ**. (1) 50 K stage with an optical IR opaque window. (2) 4K stage with copper thermalization touching the lead screw. (3) Illumination beam splitter at 1 K stage. (4) 1 K stage with a thermalized brass bearing. (5) A flexible connection in the lead screw. (6) Part of the lead screw. (7) Reinforcement copper plate. (8) The objective lens in adjustable lens tube. (9) Two $90^0$ adjustable mirrors. (10) The strain clamp with the sample illuminated. Even with all the probes attached the base temperature of <25 mK can be reached (illumination off).*

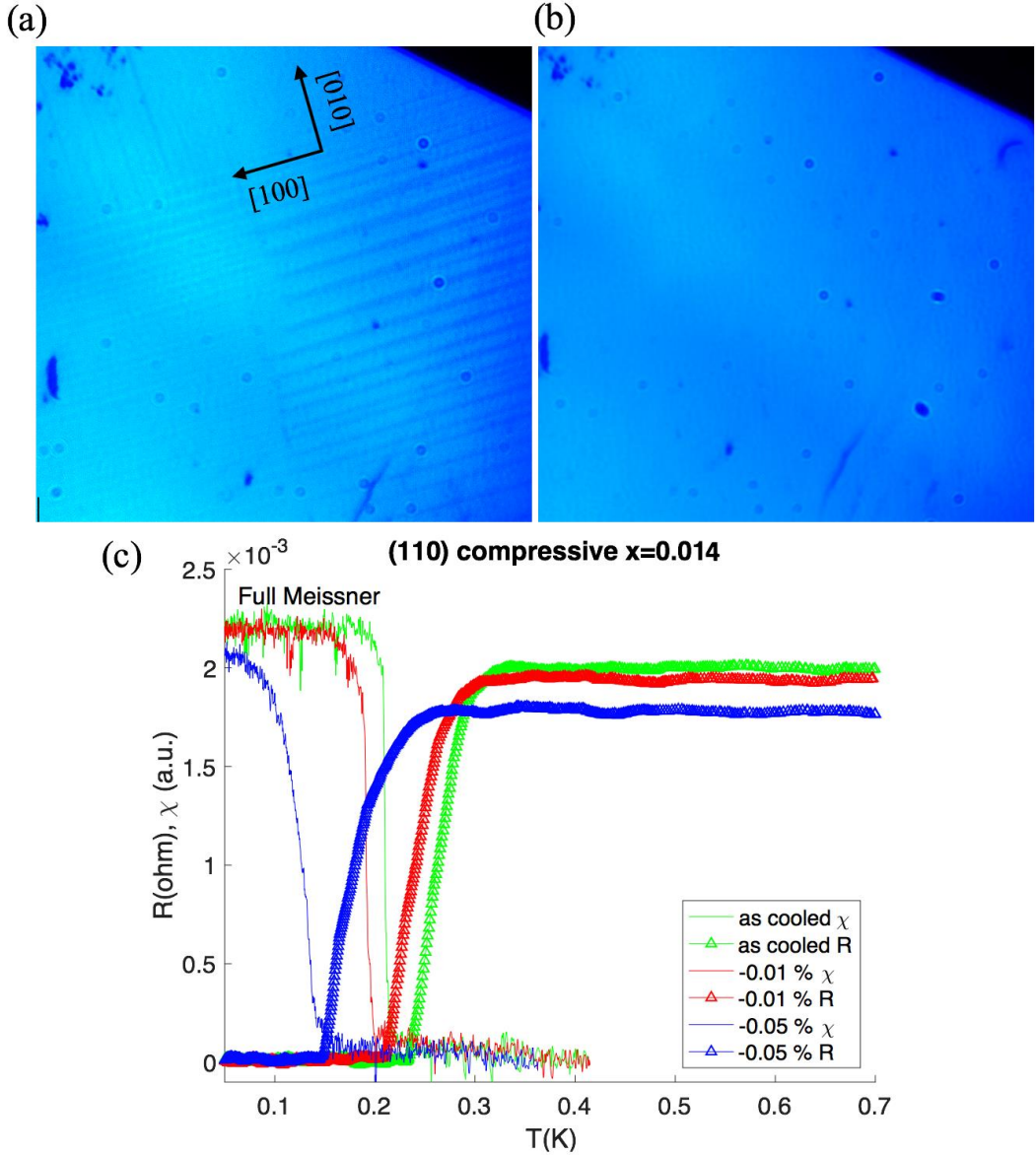

*Figure 3. **Simultaneously acquired optical images (a, b) as well as magnetic and resistive signatures (c) of the superconducting transition in an x=0.014 sample oriented with $(110)_c$ face normal to the stress direction.** The long 6-mm side (and the visible edge) of the sample was parallel to $[110]_c$. The sample was 0.5 mm thick along $[001]_c$ and 2 mm wide along $[\bar{1}\bar{1}0]_c$. Samples detwinned under $(110)_c$ compression with the c-axis aligning out of the images plane in the $[001]_c$ direction. (a) shows an as-cooled sample's surface at 40 K. The visible edge is parallel to the (110) stress. (b) shows a detwinned sample at -0.015% $[110]_c$ strain. Imaged area: 1.5x1.5 $mm^2$. The superconducting transition is strongly suppressed, and the microscopy allows us to determine the orientation of the antiferrodistortive axis which is essential to understand the effect.*

**Summary and Conclusions**

The apparatus presented here opens a new page in investigations of the interplay between strain and quantum orders. Our typical strontium titanate data set shows a reliable testing platform for behaviors across various tuning parameters and signatures of quantum orders. Several orders—superconducting, structural, polar, and magnetic can be detected. Further work includes installation of a scanning Superconducting Quantum Interference Device (SQUID) microscope at the strain clamp. [17], [18]